\documentclass[conference]{IEEEtran}
\IEEEoverridecommandlockouts 

\usepackage{cite}
\usepackage{amsmath,amssymb,amsfonts}
\usepackage{algorithmic}
\usepackage{orcidlink}
\usepackage{graphicx}
\usepackage{textcomp}
\usepackage{xcolor}
\def\BibTeX{{\rm B\kern-.05em{\sc i\kern-.025em b}\kern-.08em
    T\kern-.1667em\lower.7ex\hbox{E}\kern-.125emX}}

\usepackage{mathtools}
\DeclarePairedDelimiter\ket{\lvert}{\rangle}
\usepackage{hyperref}
\usepackage{booktabs}

\begin{document}
\bstctlcite{IEEEexample:BSTcontrol}

\title{Quantum Approximate Optimisation Algorithm for Protein Sidechain Packing}

\author{Sebastian O. M. Stewart$^{1}$\orcidlink{0009-0006-5630-7249}, Nick Chancellor$^1$\orcidlink{0000-0002-1293-0761}, Jonte R Hance$^1$\orcidlink{0000-0001-8587-7618}, and Ittoop Vergheese Puthoor$^1$\orcidlink{0000-0003-0649-5924}
\thanks{\textit{$^1$Quantum Group, School of Computing, Newcastle University, 1 Science Square, Newcastle upon Tyne, NE4 5TG, UK}}}

\maketitle

\begin{abstract}
Sidechain packing is a critical stage in protein folding, with direct implications for structure-based drug discovery. 
Google DeepMind's tool, AlphaFold2, predicts protein backbone reliably but sidechain positioning less accurately. Recovering the lowest-energy rotamer assignment over a fixed backbone is NP-hard. 
We present a hybrid quantum-classical pipeline that repacks sidechains on the AlphaFold backbone using the Quantum Approximate Optimisation Algorithm (QAOA), encoding the one- and two-body energies as a quadratic unconstrained binary optimisation (QUBO) problem. 
We introduce a constraint-preserving ansatz, pairing a W-state initialisation with a cyclic XY ring mixer, that enforces one-hot rotamer validity without penalty terms while keeping two-qubit gate scaling linear in the rotamer count. 
We also define an asymptotic shot-scaling metric, measured against the experimentally resolved conformation, 
that fixes optimiser quality independently of the baseline; its fitted growth stays below the classical exhaustive-search rate at moderate rotamer flexibility. 
Evaluated on bovine pancreatic trypsin inhibitor (5PTI) across high- and moderate AlphaFold-confidence regions, the pipeline lowers conformational energy against the AlphaFold baseline.
\end{abstract}


\section{Introduction}\label{sec:intro}
Proteins drive the mechanical and chemical processes of biological systems. A protein is fundamentally a chain of amino acids that folds into a stable, low-energy, three-dimensional geometry known as its native state --- the state most commonly found in nature. 
The simulation of this process is widely known as the protein folding problem~\cite{dill:2008}.
Experimental determination of protein structures is resource-intensive and time-consuming. In contrast, sequencing amino acids is highly efficient. 
Since a sequence largely determines its native state, efficient sequencing yields abundant inputs for which only the structure remains unknown.
If accurate structural predictions could be made computationally, researchers could efficiently identify target binding sites --- which are defined by the three-dimensional arrangement of these amino acids 

Sidechain packing is a sub-problem of protein folding that analyses how sidechains --- the chemically variable groups branching off the repeating main chain, or backbone --- fold along their dihedral \(\chi\) angles. Once linked into a chain, each amino acid unit --- a backbone segment with its attached sidechain --- is termed a \emph{residue}.
AlphaFold2~\cite{jumper:2021}, a tool developed by Google DeepMind, utilises previously resolved protein structures to determine the structure of new, unseen amino acid chains; however, while reliably generating backbone coordinates, its sidechain positioning deviates from experimental data by up to 20\%~\cite{terwilliger:2024}. 
Spatial variations in binding-site rotamers (discrete sidechain conformations) degrade molecular docking algorithms and computational compound filtering~\cite{scardino:2023}.

Sidechain packing has been proven to be NP-hard~\cite{akutsu:1997}, and quantum optimisation methods provide a possible alternative to classical heuristics: 
encoding sidechain energy minimisation as a constrained optimisation problem makes it tractable via variational quantum algorithms, specifically the Quantum Approximate Optimisation Algorithm (QAOA).

In this paper, we propose a multistep pipeline that addresses sidechain packing by building on the reliable backbone predictions provided by AlphaFold to recompute sidechain positions through QAOA. 
The proposed pipeline encodes one-body and two-body energies as a Quadratic Unconstrained Binary Optimisation (QUBO) problem. 
To enforce one-hot (exactly one rotamer selected per residue) validity without penalty terms, we introduce a constraint-preserving ansatz that combines a W-state initialisation with an $XY$ ring mixer, confining the quantum state evolution to a feasible sub-space, such as a specific Hamming-weight sector. This allows us to implement valid one-hot assignments (Hamming weight 1) without expanding the search space.
Separately, the cyclic ring topology of this mixer, in place of a fully connected one, reduces the two-qubit gate scaling. 
We show this ring-mixer construction reduces the two-qubit gate count from \(\mathcal{O}(MN^2)\) to \(\mathcal{O}(MN)\) on $MN$ qubits ($M$ residues, $N$ average rotamers per residue). Gate depth is correspondingly fixed at 2, compared with the \(\mathcal{O}(N)\) scaling of the fully connected mixer.
Finally, we construct two metrics: a shot-scaling metric (\(S(N)\)) to evaluate QAOA performance, and an energy difference (\(\Delta E\)) to compare the resulting conformations against the AlphaFold baseline structures.

\section{Background and Problem Formulation}
Under a fixed backbone, sidechain packing reduces to selecting one rotamer per residue from a discrete library so as to minimise the total conformational energy. 
With that assumption, this energy is pairwise decomposable~\cite{simoncini:2015}:
\begin{equation}\label{eq:pairwise-energy}
    E(c) = E_t + \sum_{i} E(i_r) + \sum_{i<j} E(i_r, j_s)
\end{equation}
where \(E_t\) is the backbone-dependent template energy, \(E(i_r)\) the internal and rotamer-backbone energy of rotamer \(r\) at residue \(i\), 
and \(E(i_r, j_s)\) the interaction between rotamer \(r\) at residue \(i\) and rotamer \(s\) at residue \(j\).
Minimising \(E(c)\) over all valid one-rotamer-per-residue assignments is the NP-hard combinatorial problem this pipeline targets.

The backbone is taken from AlphaFold2~\cite{jumper:2021}, which assigns each residue a pLDDT (predicted Local Distance Difference Test) score: its own confidence in the prediction.
AlphaFold reserves strict atomic accuracy, including the sidechain placement required at binding sites, for residues above 90~pLDDT~\cite{AlphaFoldProteinStructure}; at moderate confidence (pLDDT \(\approx 80\)) the backbone remains reliable while sidechain placement weakens~\cite{terwilliger:2024}. 
The fixed-backbone reduction therefore holds in both regimes.
The sidechains do not carry the same guarantee: 
they set the baseline our repacking is scored against, so that baseline is only as reliable as AlphaFold's placement, which the pLDDT itself reports.

Representing this minimisation as a QUBO, or equivalently an Ising model~\cite{lucas:2014}, expresses the energy as the Pauli-$Z$ Hamiltonian QAOA~\cite{farhi:2014} optimises.
The cost Hamiltonian encodes the energy and a mixing Hamiltonian explores the assignment space. Over \(p\) alternating Hamiltonian layers a classical optimiser tunes both until measurement probability concentrates on the low-energy states.
We adopt this framework, encoding \(E(c)\) as the cost Hamiltonian and adapting the mixer to preserve the one-hot rotamer constraint, as developed in \autoref{sec:methods}.

\section{Methodology}\label{sec:methods}
\subsection{Pipeline Overview \& Fixed-Backbone Reduction}
As established in \autoref{sec:intro}, sidechain packing is NP-hard~\cite{akutsu:1997}; classical preprocessing does not significantly reduce the number of low-energy basins
within the landscape.  
Rosetta's packer, for example, employs a stochastic Monte Carlo search over a discrete rotamer space and does not guarantee
the globally lowest-energy solution; it is therefore heuristic rather than an exhaustive optimiser.

This makes sidechain packing a natural candidate for a QAOA-based optimisation layer over a fixed backbone and rotamer library, the combinatorial problem class QAOA was introduced to address~\cite{farhi:2014}.

We propose a multistep pipeline in which a subsection of a protein is first analysed and preprocessed with PyRosetta,
followed by QAOA optimisation and finally post-processing. The first step uses PyRosetta's \verb|TaskFactory| and
\verb|ResidueLevelTask|, applying either \verb|prevent_repacking| or \verb|restrict_to_repacking| to each residue, to
fix the backbone and define flexible subsections within the protein. By fixing the backbone, the task is reduced to 
the NP-hard rotamer selection described above.

The \verb|TaskFactory| is then used alongside the \verb|ExtraRotamersGeneric| protocol to generate a large set of valid rotamer conformations positioned around the modal \(\chi \)-angle wells of the Dunbrack rotamer libraries~\cite{dunbrack:2002}. 
We disable the \verb|IncludeCurrent| task operation, which would otherwise add the input conformation to the rotamer set. 
Disabling it keeps the AlphaFold-predicted structure out of the candidate pool; however, the candidate well remains represented, so a near-optimal conformation is still 
reachable. This prevents the optimiser from returning the exact starting conformation and simply echoing the input.

Using PyRosetta's default full-atom score function~\cite{alford:2017}, one-body and two-body energies are generated for this set.
The set is then pruned by round-robin selection: rotamers are grouped by their dihedral angles (\(\chi \)) and ordered within each group by one-body energy, and the next-lowest-energy conformation is drawn from each group in turn until a pre-determined rotamer limit \(R\) is reached. \(R\) caps the retained rotamers per residue, and therefore bounds \(N\) and the qubit budget.

\subsection{QUBO Formulation of Sidechain Energies}
To execute the combinatorial search via QAOA, the pairwise energy of \autoref{eq:pairwise-energy} was formulated as a QUBO problem. 
As demonstrated by Lucas~\cite{lucas:2014}, QUBO formulations, allow classical data to be encoded into the Pauli-Z Operators (\(Z\)) required by the quantum circuit, leading to the cost Hamiltonian ($H_c$) as,

\begin{equation} \label{eq:QUBO-Formulation}
    H_C = \sum_{k \in \mathcal{R}}^{} W_k \frac{I - Z_k}{2} + \sum_{(k, l) \in \mathcal{P}}^{} W_{kl} \frac{I - Z_k}{2} \otimes\frac{I - Z_l}{2}
\end{equation}

The expression \(\frac{I - Z_k}{2}\) (and identically for $Z_l$) maps the classical binary decision variables \(c \in \{0, 1\}\) to the quantum computational basis, by shifting the eigenvalues of the Pauli-Z operator such that an active rotamer state \(\ket{1}\) yields a value of 1 (activating the weight \(W_n\)), whereas an inactive state \(\ket{0}\) evaluates to 0, nullifying the term.

One-body energy terms \(W_k\) are governed by the set of all rotamer states \(\mathcal{R}\), whereas two-body terms \(W_{kl}\) are defined by the set of interacting pairs \(\mathcal{P}\).

For a protein of \(M\) residues carrying up to \(N\) candidate rotamers each, the one-hot schema allocates \(MN\) qubits in total.
The cost Hamiltonian scales quadratically in this total rotamer count, even though the qubit count itself grows only linearly with it.
In a fully connected, worst-case interaction graph, the pairwise summation over \(\mathcal{P}\) generates \(\mathcal{O}((MN)^2)\) distinct two-qubit Pauli-Z tensor products (\(Z_k \otimes Z_l\)). In practice this bound is not reached.
PyRosetta applies an internal distance cutoff, setting interaction energies between distant residues to 0. 
The interaction matrix \(W_{kl}\) is therefore highly sparse, and the number of active terms in the Hamiltonian 
scales closer to \(\mathcal{O}(MN \cdot d)\), where \(d\) is the average number of spatial neighbours per rotamer. 

To map discrete biological rotamers to qubits, a one-hot encoding schema was selected, allocating one qubit per valid rotamer retained after classical pruning so that the qubit count grows linearly with the number of rotamers. 
Binary and Gray encodings, as in Sawaya \textit{et al.}~\cite{sawaya:2023}, reduce the qubit count for the same state space but require substantially deeper circuits. 
Domain-wall encoding~\cite{chancellor:2019} achieves a comparable constraint structure using \(N-1\) qubits and remains a viable alternative. As the circuit runtime in classical state vector simulation was the major bottleneck, one-hot encoding was adopted here.

\subsection{Constraint-Preserving Ansatz}
Initial implementations of the QAOA using a generalised Hadamard mixer layer and penalty weights to enforce the one-hot schema resulted in a high proportion (\(\sim27\% \)) of invalid bit-strings. Rather than applying harsh penalty terms within the cost Hamiltonian \(H_C\), we adopt a modified mixer Hamiltonian,
following Mancilla \textit{et al}~\cite{mancilla:2026}:
\begin{equation}
    H_{XY} = \sum_{(i, j)\in \mathcal{A}} (X_i X_j + Y_i Y_j)
\end{equation}

Here \(\mathcal{A}\) denotes the set of rotamer pairs (i,j) within a single amino acid sub-register. 
Initially, this mixing operation was structured as a complete graph, applying the \(XY\) interaction between every possible pair of rotamers within a residue. 
This dense topology yields a quadratic gate scaling of \(\mathcal{O}(M \frac{N (N-1)}{2}) = \mathcal{O}(M N^2)\), where \(M\) is the number of residues analysed and \(N\) is the average number of valid rotamer conformations per residue.

\begin{figure}[htbp]
\centerline{\includegraphics[width=0.88\columnwidth]{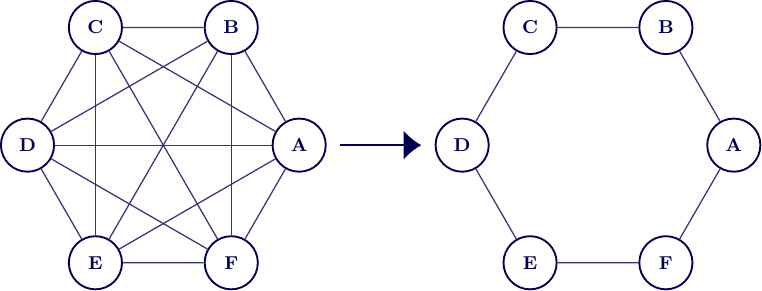}}
\caption{
    Transition from the complete-graph mixer to the cyclic-ring mixer within a residue sub-register, reducing the two-qubit gate count from \(\mathcal{O}(MN^2)\) to \(\mathcal{O}(MN)\) while preserving full reachability of the valid one-hot subspace across QAOA layers.
}\label{fig:complete-to-ring-mixer}
\end{figure}

To mitigate this overhead, the mixing topology was redefined as a one-dimensional cyclic graph (see \autoref{fig:complete-to-ring-mixer}).
By restricting the Pauli-X (\(X\)) and Pauli-Y (\(Y\)) Operators strictly to the two adjacent rotamers in a closed loop, the scaling reduces to \(\mathcal{O}(MN)\). 
Because the ring remains a \emph{connected} graph, the mixer stays irreducible over the valid subspace, so the full state space remains reachable across multiple QAOA layers despite the compressed circuit depth. 

The XY mixer conserves Hamming weight: the \(X_i X_j + Y_i Y_j\) coupling hops a single excitation between adjacent rotamers without changing the total excitation count.
This allows us to drop the penalty term, but imposes a pre-condition. 
Because the mixer never leaves the subspace it starts in, it enforces one-hot validity only if the optimiser begins inside the Hamming-weight-1 subspace; started from an invalid configuration, it would preserve that invalidity just as faithfully. 
We therefore initialise each residue sub-register in a W state, the equal superposition over all single-excitation basis states:
\begin{equation}\label{eq:w-state}
    \ket{W_n} = \frac{1}{\sqrt{n}} \sum_{i=1}^{n}\ket{e_i}
\end{equation}
where \(\ket{e_i}\) denotes the computational basis state in which only the \(i\)-th rotamer is active.
Initialisation was implemented using PennyLane's \verb|StatePrep| operation. 
While computationally efficient for classical state vector simulations, this operation abstracts underlying gate-level mechanics. 
Consequently, deployment on quantum hardware would require decomposing this step into hardware-native W state preparation circuits.

By initialising the system in this W state, the algorithm begins entirely within the valid Hamming-weight-1 biological subspace, and the ring mixer holds it there across every subsequent layer.
This pairing enforces the one-hot constraint structurally, removing the penalty term that earlier produced the \(\sim27\% \) invalid bit strings. 
The resulting QAOA layer combines this initialisation, the global cost Hamiltonian, and the localised ring mixer.

\subsection{Warm-Starting \& Parameter Caching}
To evaluate the QAOA across varying circuit depths (\(p\)), an initial cold-start methodology was established, wherein parameters for each depth \(p \in \{1, 2, 4, 8, 12\} \) were initialised independently, discarding the prior results.

Following initial testing, this was replaced by a warm-starting model, as several cold-start runs stalled rather than improving with depth.
In a five-residue subsection (residues 18--22 on the 5PTI protein identifier for Bovine Pancreatic Trypsin Inhibitor~\cite{wlodawer:1984}), averaged over 30 random seeds, the cold-starting model could not break past a plateau at approximately 20\% target probability. 

Rather than independently initialising parameters for deeper circuits, the previous best optimised parameters \(\gamma_{1:p}\) and \(\beta_{1:p}\) were saved and reused.
The new layers were then zero-padded:
\begin{equation}
    \gamma_{p+1} = \beta_{p+1} = 0
\end{equation}

Initialising the new layers in this fashion placed the optimiser at a saddle point. Following the pretrubed gradient method~\cite{jin:2017}, saddle points can be escaped efficiently by adding a small amount of Gaussian noise (\(\epsilon \sim \mathcal{N}(0, 10^{-4})\)) to the newly-added parameters to break the symmetry. 
Gradients were computed by adjoint differentiation~\cite{jones:2020}, a single backward pass per step rather than the two circuit evaluations per parameter required by parameter-shift, which kept the deeper-circuit optimisations tractable.

\subsection{S(N) asymptotic shot-scaling metric}\label{sec:snmetric}
To evaluate QAOA performance, we established the metric asymptotic shot-scaling limit (\(S(N)\)). 
This required defining a `success state' within the rotamer space.
While the success criterion was initially restricted to the Global Minimum-Energy Conformation (GMEC) --- identified via exhaustive search --- the definition of success was expanded to reflect biological reality.
In nature, proteins often exist in a structural ensemble due to variations in experimental conditions such as solvent composition or temperature.
Defining the GMEC over the enumerated rotamer combinations of the ground-truth structure lets $S(N)$ measure the optimiser's ability to locate the best combination within a given candidate set, independent of whether that set is itself biologically optimal.

Following prior research~\cite{lowegard:2020}, which applied a window of 2~kcal/mol from the GMEC, a more conservative value of 1.5~kcal/mol was adopted for this work. 
This tighter limit ensures that `success' is only attributed to configurations that are energetically close to the optimal packing.
Because PyRosetta's score function operates on a kcal/mol scale, this threshold was applied directly to the Hamiltonian expectation values.

We analysed the final probability distribution of the QAOA by summing the probabilities of all success-states to yield a cumulative success probability \(P_c(N)\), where \(N\) denotes the total rotamer count of the analysed subsection: 
\begin{equation}
    P_{c}(N) = \sum_{n \in \mathcal{S}} P_{n}
\end{equation}
To translate this theoretical metric to a practical metric, we converted \(P_{c}\) to a shot-scaling metric:
\begin{equation}\label{eq:shot-scaling-metric}
    S(N) = \left\lceil \frac{\log(1 - \alpha)}{\log(1 - P_{c}(N))} \right\rceil
\end{equation}
where \(\alpha \) represents the required recovery probability (\(99.99\% \)).
Since the state space of explorable rotamer conformations scales exponentially, determining the GMEC becomes impossible at higher qubit counts.
Therefore, it is instead calculated at tractable sizes and extrapolated, giving the sampling overhead required to recover a viable biological solution at scale.

\section{Results}
\subsection{Experimental Setup}\label{sec:setup} 
The pipeline described in \autoref{sec:methods} was run on the structural file of the 5PTI protein, with the AlphaFold prediction (AF-P00974-F1-v6)~\cite{jumper:2021, bertoni:2026}, using PyRosetta~\cite{chaudhury:2010} (2026.03 quarterly release, build 5e498f1\footnote{PyRosetta4.Release.python313.m1 2026.03+releasequarterly.5e498f1409c68ade56c8ce5842bf79e1b02e8db4}). 
The two structures serve distinct roles: the experimental PDB structure provides the ground-truth rotamer space from which the GMEC, and hence the shot budget of \autoref{sec:snmetric}, is determined by exhaustive search.
The AlphaFold prediction serves as the baseline for the second stage. 
Once $S(N)$ has been determined, the QAOA is run on the AlphaFold backbone, and the AlphaFold energy is reused as the reference against which the resulting rotamer combination is scored.

PennyLane 0.44.1~\cite{bergholm:2018} was used alongside PennyLane Lightning 0.44.0 and JAX 0.7.2 to simulate and optimise the QAOA circuit, executed on Google Colab A100 GPUs. 
To account for the stochasticity of the Adam optimiser in Optax 0.2.8, each analysed subsection of the 5PTI protein was run over 30 random seeds (1--30), a heuristic choice to ensure stable summary statistics.
The per-residue rotamer cap \(R\) ranged from 3 to 7, set per subsection to keep the qubit count tractable.

To compare the resulting rotamer conformations against the AlphaFold baseline, we chose the classification metric \(\Delta E = E_{QAOA} - E_{AF}\). 
Here, \(E_{AF}\) denotes the energy score PyRosetta assigns to the baseline AlphaFold protein, whereas \(E_{QAOA}\) denotes PyRosetta's energy score after the results are extracted from the QAOA. 
As before, we used the 1.5~kcal/mol threshold, applied here as a symmetric band about the AlphaFold baseline rather than the one-sided window from the GMEC in \autoref{sec:snmetric}. Biologically, this is to separate meaningful energy differences from insignificant ones, and thereby creating three classifications: 
\begin{itemize}
    \item Class I (Wins) for \(\Delta E < -1.5~\text{kcal/mol}\),
    \item Class II (Ties) for \(|\Delta E| \leq 1.5~\text{kcal/mol}\),
    \item Class III (Losses) for \(\Delta E > 1.5~\text{kcal/mol}\).
\end{itemize}

Full problem instances and the scripts that reproduce them are provided in the \hyperref[sec:code]{Code and Data Availability} release.

\subsection{Shot-scaling validation, S(N)}\label{sec:shot-scaling-results}
\begin{figure}[htbp]
    \centerline{\includegraphics[width=0.88\columnwidth]{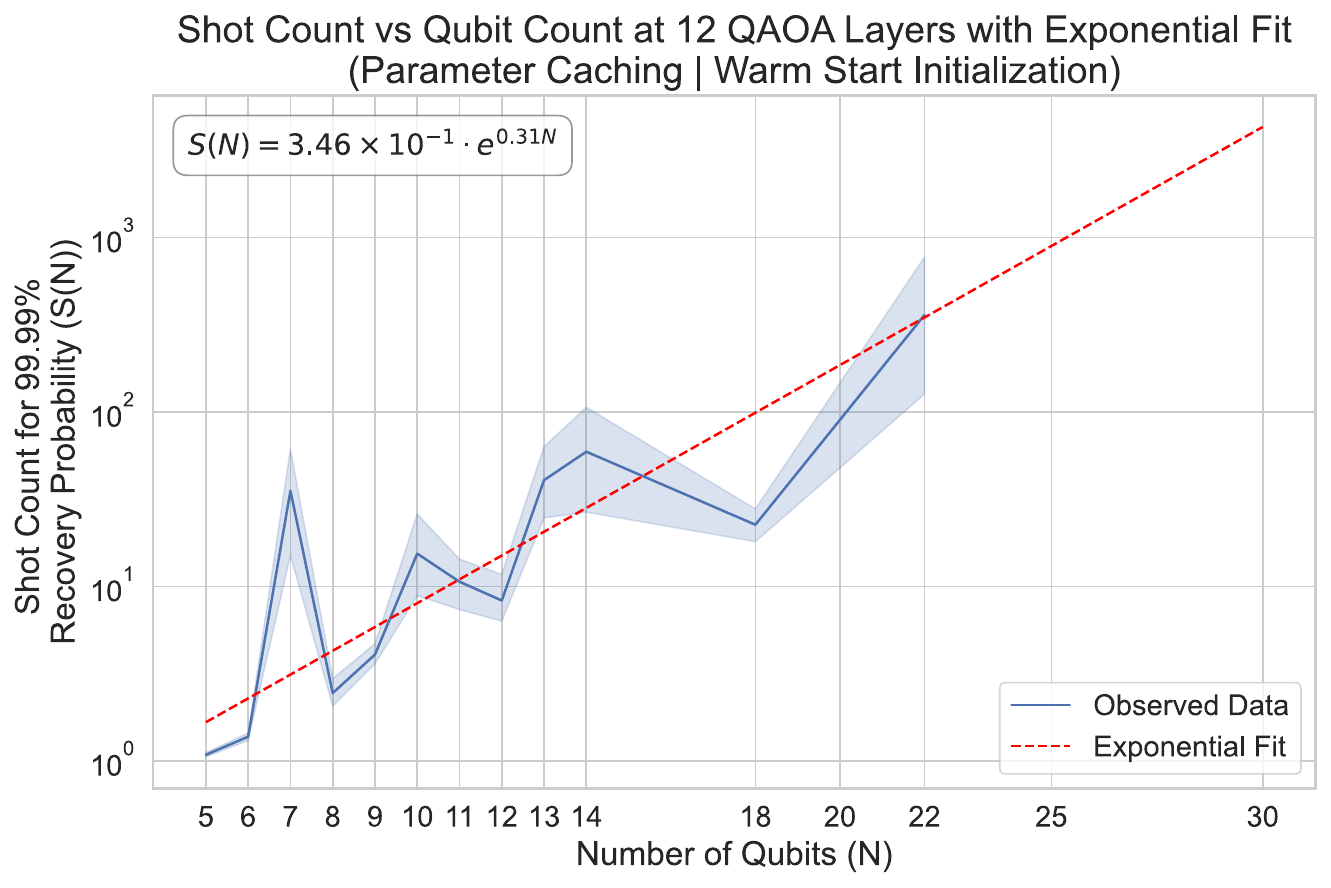}}
    \caption{
        Asymptotic shot scaling \(S(N)\) at circuit depth \(p=12\) on the 5PTI protein, for a \(99.99\% \) recovery probability, pooled across all evaluated residue subsections and their 30 seeds at each qubit count. The shaded area is a bootstrapped 95\% CI over this distribution.
        Shot count appears to grow exponentially in qubit count; the fitted model stays below \(500\) shots through \(N=22\). 
        \(N=7\) is an outlier; the bands widen at the sparsely sampled large-\(N\) points. 
    }\label{fig:shot-scaling-limit}
\end{figure}

We measured \(S(N)\) on the ground-truth 5PTI protein set that is established in \autoref{sec:setup} at the maximum circuit depth \(p=12\) across the qubit spectrum (\autoref{fig:shot-scaling-limit}), under the declared one-sided \(1.5\)~kcal/mol success window for a \(99.99\% \) recovery probability. The overall trend is the expected exponential growth, but the distribution is non-monotonic: \(N=7\) is an outlier that spikes above the fit before collapsing at \(N=8\), and the sampled band widens at the larger, sparsely sampled qubit counts (\(N=18\)--\(22\)). 
This spread indicates that the required probability mass is sensitive to the specific localised Hamiltonian (driven by the energy landscape of the sampled 5PTI rotamers) rather than scaling purely with Hilbert-space dimension. 
If convergence were dictated solely by search-space size, shots would scale monotonically with qubit count; they do not.
Re-indexing the same data by residue subsection starting index at fixed qubit count shows a strong dependence on which residues are sampled, (shown in \autoref{tab:topology}). 

To set a conservative upper bound for the second-stage sampling budget, we fit the exponential regression shown in \autoref{fig:shot-scaling-limit}. 
Despite the exponential scaling, the budget stays viable for high-performance computing: the fit remains below \(500\) shots through 22 qubits. 
This follows from the one-hot encoding, under which the quantum dimension grows with the sum of available rotamers (\(\sum R_i = N\)), while the classical search space grows with their product (\(\prod R_i\)); the algorithmic advantage therefore depends on how the rotamers are distributed across the target residues. 

\begin{table}[htbp]
    \caption{Shot count versus qubit count (\(p=12\), \(99.99\% \) recovery, 5PTI protein). For each residue subsection we take the mean shot count over 30 seeds; a given qubit count \(N\) is shared by several subsections, since equal qubit counts can arise from different residue subsections with different budgets. Min and Max report the lowest- and highest-budget subsections at \(N\), illustrating that \(N\) alone does not determine cost.
    }\label{tab:topology}
    \centering
    \begin{tabular}{rrrr}\toprule
    Qubits \(N\) & Min shots & Max shots & Max/Min \\
    \midrule
    7  & \(\sim 1.10\) & \(\sim 84.27\)  & \(\sim 77\times \) \\
    12 & \(\sim 3.40\)  & \(\sim 24.80\)  & \(\sim 7\times \) \\
    22 & \(\sim 52.53\) & \(\sim 732.88\) & \(\sim 14\times \) \\\bottomrule
    \end{tabular}
\end{table}

\subsection{High AlphaFold-confidence regions, pLDDT \textgreater{} 90}
Using the sampling budget fixed by \(S(N)\) (\autoref{sec:snmetric}), the QAOA repacks the AlphaFold backbone in regions of high predicted confidence (specifically, AlphaFold's per-residue pLDDT). Each run is classified by \(\Delta E = E_{QAOA} - E_{AF}\) under the symmetric \(1.5\)~kcal/mol band of \autoref{sec:setup}.

\begin{figure}[htbp]
    \centerline{\includegraphics[width=0.88\columnwidth]{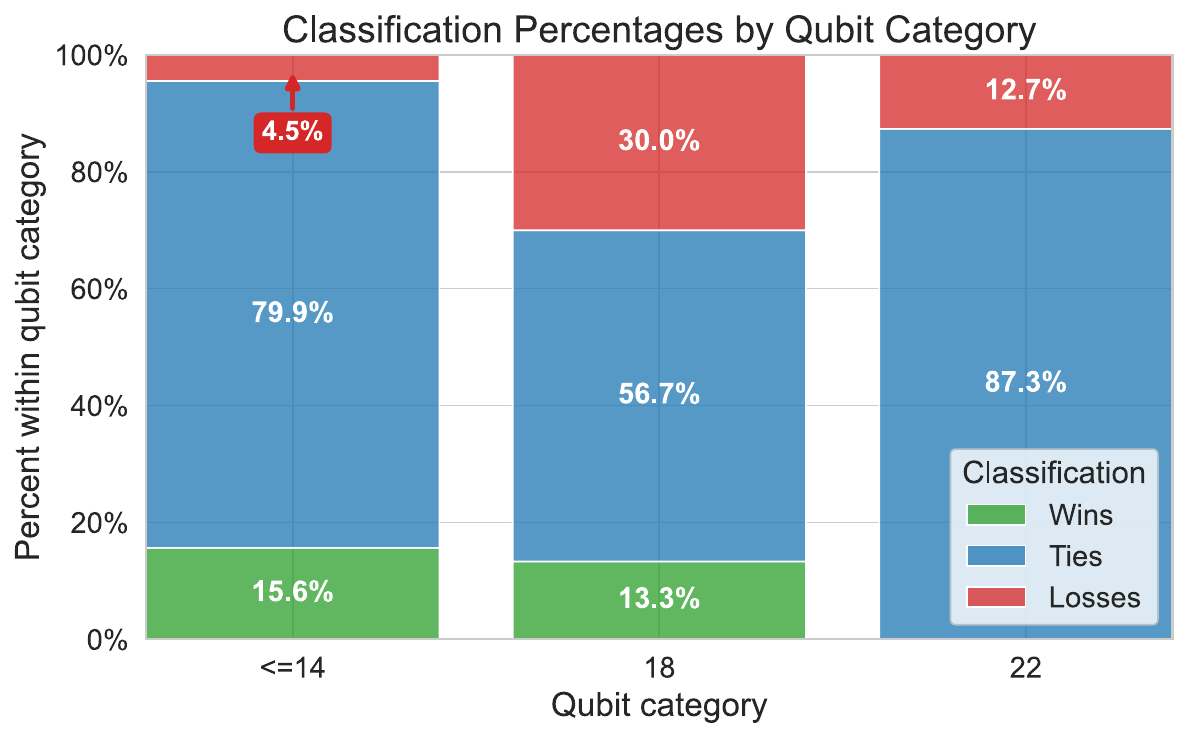}}
    \caption{
        Classification outcomes by qubit category, high AlphaFold-confidence regions (pLDDT \textgreater{} 90),        \(\Delta E\) against the AlphaFold baseline. Wins decline from \(15.6\% \) (\(\le 14\) qubits) to zero at 22. Losses peak in frequency at 18 qubits (\(30.0\% \), penalties up to \(+3.973\)~kcal/mol) but in magnitude at 22 (\(12.7\% \) of runs, up to \(+4.393\)~kcal/mol), where outcomes are otherwise \(87.3\% \) ties. 
    }\label{fig:hiconf-scaling}
\end{figure}

At register sizes (5--14 qubits) the optimiser wins on \(15.6\% \) of runs, ties on \(79.9\% \), and loses on \(4.5\% \) (\autoref{fig:hiconf-scaling}). 
The aggregate win rate understates the effect, because wins are not spread across the register: they concentrate almost entirely in the residue 57--64 window, where per-subsection win rates approach unity (e.g.\ 30/30 at residues 60--63), while neighbouring sub-registers of equal qubit count return ties. 
As with \(S(N)\), performance is governed by which residues are sampled, not how many qubits encode them; aggregated across runs, \(\Delta E\) correlates moderately with residue length (\(r \approx -0.43\), \(p < 0.001\)) while raw qubit count has minimal stand-alone effect. 
Across the 230 winning runs the mean improvement is \(-1.864\)~kcal/mol, with a peak of \(-2.29\)~kcal/mol. 

Extending to 18 and 22 qubits removes this advantage. 
At 18 qubits wins fall to \(13.3\% \) and losses spike to \(30.0\% \), with the discretisation penalty (i.e., penalty due to sidechains being positioned too close to one another) in the failing cases reaching \(+3.973\)~kcal/mol. 
At 22 qubits no run wins, ties dominate (\(87.3\% \)), and while losses are less frequent (\(12.7\% \)), they reach a larger maximum penalty of \(+4.393\)~kcal/mol. 

Degradation is not a solver failure. 
Because \(S(N)\) is measured against the experimentally resolved structure, optimiser quality is established independently of baseline quality: the QAOA concentrates probability on the optimum within a given candidate set. 
In high-pLDDT regions, the AlphaFold baseline is continuous and already near-optimal.
The Dunbrack library is finite, so forcing a wider span of these already-good residues into it can only raise the energy of the best available discrete conformation.
That conformation therefore drifts up toward the baseline, producing ties; where it overshoots, the run is logged as a loss ---
more often at 18 qubits, and by a larger margin at 22. 
The optimiser correctly identifies this discrete optimum; the resulting ties and losses reflect the cost of discretisation at scale, not a limit of the quantum optimiser.

\subsection{Moderate AlphaFold-confidence regions, pLDDT \texorpdfstring{\(\approx \)}{} 80}
To test the pipeline's generalisation, we extended evaluation from high-confidence domains (pLDDT \(>90\)) to moderate-confidence regions (pLDDT \(\approx 80\)). 
AlphaFold's own validation reserves strict atomic accuracy, including the sidechain placement required at binding sites, for residues above \(90\)~pLDDT~\cite{AlphaFoldProteinStructure}.
At \(\approx 80\)~pLDDT the backbone remains reliable while sidechain packing is weaker.
Since the fixed-backbone assumption still holds but \(E_{AF}\) is a weaker baseline, we expected a higher win rate and larger \(\Delta E\).

Across 1,500 runs at 5--14 qubits, 810 (\(54\% \)) beat the baseline and only one was a Class~III loss (\(99.93\% \) no-loss). 
The improvements were also larger: a mean of \(-8.114\)~kcal/mol across the winning runs and a peak of \(-16.265\)~kcal/mol at residues 6--12. 
We attribute these larger gains to the weaker AlphaFold baseline in these regions, not to a stronger quantum solution: \(S(N)\) fixes optimiser quality against the resolved structure (\autoref{sec:snmetric}), the inverse relationship between AlphaFold confidence and \(\Delta E\) reflects baseline weakness rather than added quantum capability. 
These magnitudes therefore describe a proof-of-concept setting, bounded to 5--14 qubits by compute-credit limits, and are not validated against downstream binding affinity or structural stability.

\section{Discussion}
This work demonstrates a hybrid quantum-classical pipeline for sidechain packing in which a constraint-preserving ansatz enforces one-hot validity. 
Alternative mixer topologies reduce two-qubit gate scaling from \(\mathcal{O}(MN^2)\) to \(\mathcal{O}(MN)\), and the \(S(N)\) metric provides a predictive shot budget: the samples needed to recover the GMEC with \(99.99\%\) probability, fitted where exhaustive verification is possible and extrapolated beyond it. 
Its fitted growth rate (\(e^{0.31N}\)) stays below the classical exhaustive-search rate (\(e^{(\ln R / R)\,N}\)) for rotamer concentrations \(R \le 5\) (see \autoref{sec:shot-scaling-results}); the binding constraint is the \emph{average} density across a sub-register, not a per-residue cap, so heterogeneous topologies that mix high- and low-flexibility residues preserve the advantage even where individual residues exceed five rotamers.
For these sub-registers the budget grows strictly slower than the search space it replaces, cutting the candidates requiring classical rescoring by up to an order of magnitude at large register sizes. 
The advantage inverts only for sub-registers composed almost entirely of high-flexibility residues, such as arginine or lysine clusters.

While improvements appear against both high and moderate AlphaFold-confidence baselines, they do not by themselves establish genuine optimisation: the larger \(\Delta E\) in moderate-confidence regions tracks baseline weakness, and may therefore reflect a weak-baseline artifact rather than added quantum capability. 
Downstream validation is therefore needed, but it does not undermine the \(S(N)\) metric. 
Because \(S(N)\) is measured against the experimentally resolved GMEC, optimiser quality is established independently of baseline quality: the metric therefore guarantees optimal selection within any given rotamer candidate set regardless of that set's baseline quality.

\section{Conclusion \& Future Work}

The constraint-preserving ansatz removes penalty terms while cutting gate scaling, and the \(S(N)\) metric enables shot estimation in large-qubit simulations where classical exhaustive search becomes intractable.
Because biological fidelity is set by the preprocessing and energy model rather than by the optimiser, it is decoupled from the optimisation and can be validated as a separate problem.

Whether the (\(S(N)\)) budget generalises beyond 5PTI remains an open question, as does the effect of hardware noise on sampling 
fidelity once these methods move onto real quantum devices. Linking the energy improvements to binidng affinity would further establish structural relevance for drug discovery.

\section*{Acknowledgment}

JRH acknowledges support from a Royal Society Research Grant (RG/R1/251590), and from their EPSRC Quantum Technologies Career Acceleration Fellowship (UKRI1217). NC and JRH also acknowledge support from an EPSRC Mathematical Sciences Small Grant (UKRI3647). NC acknowledges support from QCi3 - the Hub for Quantum Computing via Integrated and Interconnected Implementations (EP/Z53318X/1).

\phantomsection
\section*{Code and Data Availability}\label{sec:code}
The code and data supporting this work are openly available at \href{https://doi.org/10.5281/zenodo.21038963}{doi.org/10.5281/zenodo.21038963}.

\bibliographystyle{IEEEtran}
\bibliography{references} 

\end{document}